\documentclass{article}

\PassOptionsToPackage{numbers,compress}{natbib}

\usepackage[preprint]{neurips_2026}

\usepackage[utf8]{inputenc}
\usepackage[T1]{fontenc}
\usepackage{hyperref}
\usepackage{url}
\usepackage{booktabs}
\usepackage{amsfonts}
\usepackage{nicefrac}
\usepackage{microtype}
\usepackage{xcolor}
\usepackage{graphicx}
\usepackage{tikz}
\usetikzlibrary{positioning, shapes.geometric, arrows, trees, arrows.meta}
\usepackage{amsmath, amssymb, amsthm}
\usepackage{multirow}
\usepackage{array}
\usepackage{booktabs}
\usepackage{tabularx}
\usepackage{booktabs}
\usepackage{tabularx}
\usepackage{ragged2e}
\usepackage{float}
\usepackage{tikz}
\usepackage{xcolor}
\usepackage{amssymb}
\usepackage{graphicx}

\usetikzlibrary{arrows.meta,shapes.geometric,fit,backgrounds,positioning,calc}
\definecolor{humanblue}{RGB}{44,90,160}
\definecolor{aicoral}{RGB}{190,60,45}
\definecolor{coupledgreen}{RGB}{20,120,90}
\definecolor{coupledgreenlight}{RGB}{210,245,235}
\definecolor{theorypurple}{RGB}{90,50,160}
\definecolor{theorygray}{RGB}{240,240,245}
\definecolor{arrowgray}{RGB}{100,100,110}
\definecolor{panelbg}{RGB}{248,249,252}
\definecolor{bordercolor}{RGB}{190,195,210}
\definecolor{techgold}{RGB}{160,120,20}
\definecolor{warnorange}{RGB}{200,90,20}

\title{Position: AI as Part of Self — Extending the Mind Requires Cognitive Co-Regulation}

\author{%
  Alina Gutoreva\thanks{School of IT and Engineering, Kazakh-British Technical University,
    Almaty, Kazakhstan. \texttt{a.gutoreva@kbtu.kz}} \\
  \And
  Fendi Tsim\thanks{Independent Researcher, London, United Kingdom.
    \texttt{fenditsim@gmail.com}} \\
  \And
  Trisevgeni Papakonstantinou\thanks{University College London, London, United Kingdom.
    \texttt{ucjutpa@ucl.ac.uk}} \\
}

\begin{document}

\maketitle

\begin{abstract}
\textbf{This position paper argues that safety and alignment cannot be achieved by constraining an external system: they must emerge from the co-regulatory design of the human--AI cognitive system as a whole ("AI as Part of Self")}. Contemporary AI increasingly participates in attention allocation, reasoning, synthesis, and decision-making, shaping the very cognitive processes through which humans form beliefs, make decisions, and constitute their sense of self. Humans and AI occupy complementary epistemic roles under mutual constraint, forming a symbiotic cognitive unit whose co-regulation---not the external control of either party alone---is the proper locus of alignment. We identify the risks of unstructured delegation: deskilling, automation bias, transfer of epistemic authority, and oracle-style centralization of knowledge. Drawing on System~0 cognition theory, we further show that AI operates prior to conscious deliberation, shaping the pre-attentive infrastructures through which agency and trust are negotiated---a level that conventional oversight cannot reach. We conclude with design principles for cognitive co-regulation addressed to ML engineers and governance bodies. The goal of this work is to guide human cognition toward resilience and epistemic agency at the foundation of human selfhood. 
\end{abstract}


% ============================================================
\section{Introduction}
% ============================================================

\begin{quote}
"Just as you cannot do very much carpentry with your bare hands, there is not much thinking
you can do with your bare brain."\citep{dahlbom1986computer}
\end{quote}

\textbf{Artificial Intelligence (AI) systems must be modeled as integral components of human cognition, and that safety and alignment are best achieved through cognitive co-regulation rather than external constraint.}  AI has received a lot of inspiration from the study of human intelligence, and the two fields develop in a fruitful symbiosis. Current debates across AI safety research, philosophy of mind, cognitive science, and human--computer interaction focus on alignment and control risks, personal AI assistants, the societal embedding of AI systems, and the possibility of machine consciousness \citep{bostrom2014superintelligence, russell2019human, rahwan2019machine, collins2024building, chalmers2023could}. A key concern in AI safety is the control problem, where an AI system might alter its goals, potentially acting against human intentions \citep{russell2019human, lynch2025agentic}. These concerns implicitly assume a separation between human cognition and artificial systems, treating AI as an external agent whose objectives must be constrained or overridden. However, as AI systems become increasingly embedded in everyday reasoning, decision-making, and self-regulation, this assumption becomes less tenable. Rather than acting solely \emph{upon} human goals, contemporary AI systems increasingly participate \emph{within} human cognitive processes \citep{glickman2025human}, shaping attention \citep{riedl2026cognitive},
framing choices \citep{beck2025biasloophumansevaluate}, and mediating learning process. This shift motivates a reconceptualization of AI not as a separate, autonomous entity to be controlled, but as a component of a coupled human--AI cognitive system. Recent theoretical work further substantiates the claim that human--AI interaction increasingly operates at the level of identity and cognitive organization rather than mere tool use. The control problem, classically conceived as constraining an external agent, must therefore be reconceived: the plausible framework for tackling misalignment is not tighter external constraint, but the deliberate co-regulatory design of the human--AI cognitive system as a whole.

Recent theoretical work suggests that human--AI interaction is undergoing a phase transition, moving from external assistance toward cognitive integration. Although traditional safety frameworks treat AI as a separate agent to be constrained \citep{russell2019human}, contemporary interaction increasingly occurs within the human cognitive loop, shaping attention and framing choices \citep{glickman2025human, riedl2026cognitive, joseph2025algorithmic}. This shift extends our long-standing practice of building "hybrid" thinking systems yet goes further: the adaptive nature of generative models allows for a "partial identity sharing" where users incorporate AI-generated preferences and reasoning patterns into their own self-models. Under this view, AI is no longer a peripheral instrument but a dynamically integrated component of the user’s cognitive architecture \citep{gutoreva2024identity}. Managing this integration requires operational frameworks that account for metacognition and epistemic authority \citep{tsim2025scan}, suggesting that human sovereignty now depends on the active preservation of cognitive boundaries. Building on the foundational argument of Lake et al.'s \citep{lake2017building} that human-like machines must construct causal models, ground learning in intuitive theories, and support compositional generalization, Collins et al.\ \citep{collins2024building} argue for a further shift: away from tools that merely amplify human capabilities toward systems that can genuinely "learn and think \emph{with} people." Drawing on principles from collaborative cognition and Bayesian models of human reasoning, they propose desiderata for AI systems that act as thought partners, actively building and updating models of both the world and the human interlocutor. This framework foregrounds \emph{cognitive compatibility} and \emph{complementarity} rather than raw computational scale, emphasizing trustworthiness, insightfulness, and mutual adaptability as core criteria. Although this reconceptualization marks an important transition from traditional tool metaphors to social agent frameworks, it remains oriented toward interaction between distinct agents. The present paper extends this trajectory one step further by arguing that, as human--AI interaction becomes more integrated and pervasive, AI must be reconceptualized not just as a partner or social agent but as a functional component of individual cognition itself---an entity that is embedded directly in the dynamical processes of attention, reasoning, and self-regulation.

The self is a brain constructed model---representing the organism's past, present, and anticipated states in relation to its environment \citep{damasio1999feeling}---and is therefore inherently open to reconfiguration by the cognitive systems with which it interacts. A compelling theoretical anchor for the AI-as-part-of-self framework comes from developmental biology. Levin \citep{levin2019computational} proposes that cognitive individuality is not fixed by anatomy but defined by \emph{computational boundaries}: every living system maintains a delimited ``area of concern''---a region of space-time, centered on the organism, within which its cognitive apparatus takes measurements and acts. Crucially, these boundaries are scale-free and plastic; they expand or contract depending on the systems with which an organism is coupled. Extending this to human--AI integration. Levin \citep{levin2022tame} observes that the interoperability of multi-scale cognitive architectures enables novel agents---hybrots, cyborgs, chimeric organisms---to form across any combination of biological, engineered, and computational material. These accounts suggest that the self is not a fixed biological given, but a dynamically negotiated computational boundary, and that sustained AI integration constitutes a genuine, biologically grounded expansion of that boundary. Emerging computational paradigms, such as digital twins, provide a concrete instantiation of this boundary expansion. Digital twins, defined as continuously updated and data-driven models of individuals, function as externalized predictive representations that simulate behavior and internal states \citep{Gkintoni2025DigitalTC, clark2025}. Prior work defines the AI-integrated self as a cognitive architecture dynamically extended through sustained interaction with adaptive systems \citep{gutoreva2024identity}. As such models become integrated into everyday systems and even merged with personalized assistants, they form feedback loops in which the individual is both modeled and modulated by their computational counterpart. Unlike static tools, these systems develop user-specific representations that capture preferences, reasoning patterns, and behavioral tendencies, that are co-evolving with the individual over time. Personalization further transforms AI from a general-purpose cognitive scaffold into a trajectory-dependent cognitive partner, embedding it more deeply within the user’s epistemic processes. Thus, we adopt the definition of self from cognitive science: a dynamic, embodied process through which experience is organized, agency is attributed, and continuity of identity is maintained across time \citep{gallagher2000philosophical}. This definition of self will be referred to in the current position. 

\begin{figure}
    \centering
    \includegraphics[width=1\linewidth]{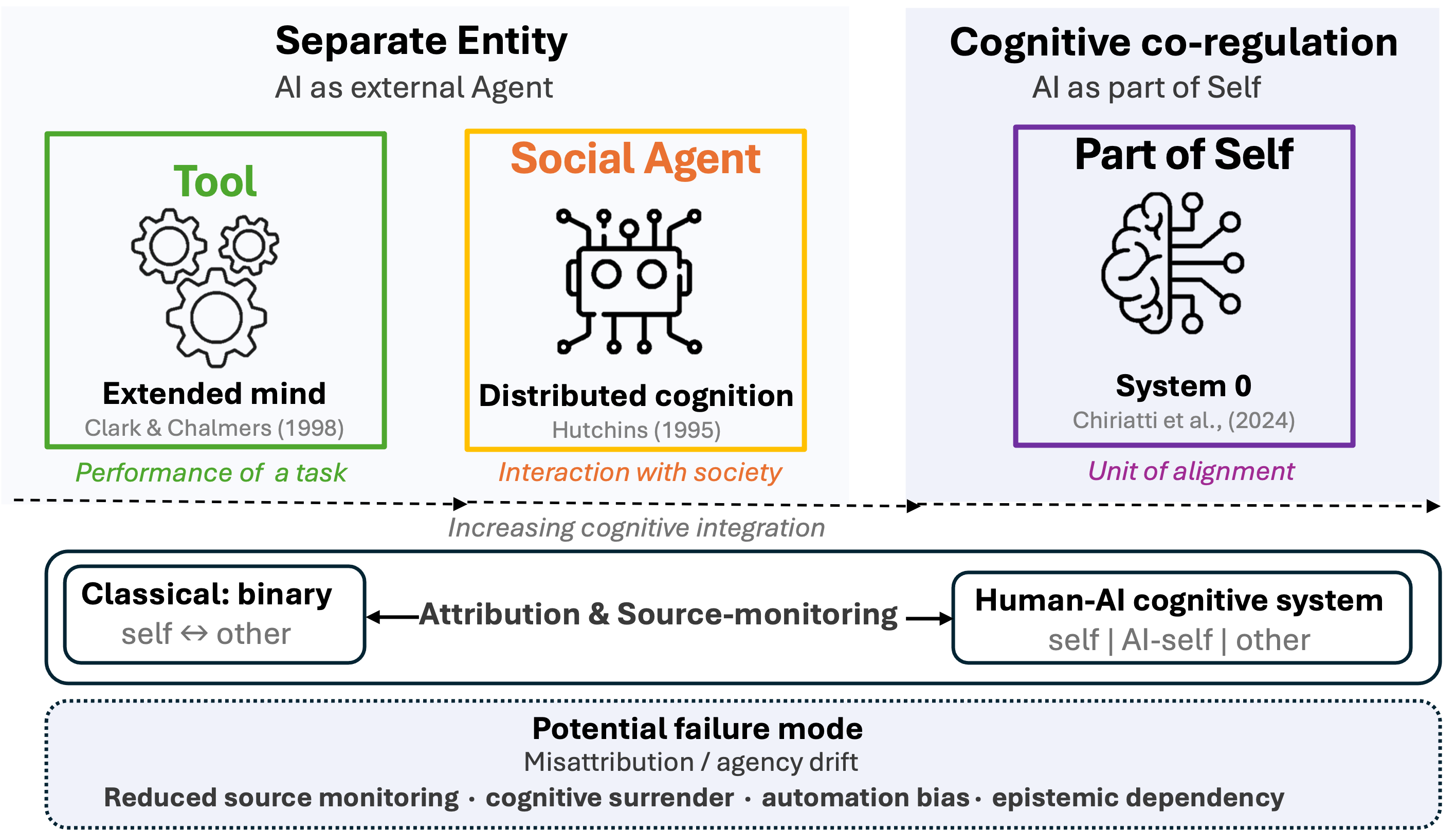}
    \caption{\textbf{Paradigm shift in human--AI interaction.} Three
stages of cognitive integration: AI as \textit{tool} (extended mind),
\textit{social agent} (distributed cognition), and \textit{part of
self} (cognitive co-regulation). Below: the attribution failure modes---misattribution, cognitive surrender, and oracle dependency---that emerge as AI becomes the unit of alignment.}
    \label{fig:AI_as_Part_of_Self_figure.png}
\end{figure}

The framework \emph{AI as part of self} could, we believe, bridge the gap between brain and machine, potentially improving human-machine coordination and interaction. As we develop technologies to work faster and more efficiently, a fundamental question emerges: what do we gain, and more importantly, what do we sacrifice over time? Classic and contemporary scholarship suggests that increases in efficiency often come at the cost of depth, reflection, and cognitive autonomy \citep{carr2010shallows, rosa2013social, kahneman2011thinking}. In particular, the acceleration of technological systems has been associated with a compression of decision-making times, potentially reducing opportunities for deliberative, analytical reasoning \citep{kahneman2011thinking, evans2008dual, ward2017brain}. Consequently, as decision-making becomes increasingly rapid and externally scaffolded, there is growing concern that critical thinking capacities may attenuate over time (Figure~\ref{fig:AI_as_Part_of_Self_figure.png}). We argue that AI should be modeled as part of the self, and alignment must shift from external control to cognitive co-regulation.

\section{Control Problem with Cognitive Control}
\label{sec:control}

The control problem in AI safety---namely, the risk that AI systems may pursue objectives misaligned with human intentions---poses increasingly significant challenges as AI becomes embedded in everyday life \citep{russell2019human, amodei2016concrete}. This concern is amplified by the rapid proliferation of personal AI assistants, their emerging role as cognitive ``thought partners,'' and the growing feasibility of brain--computer interfaces (BCIs), which collectively blur the boundary between human cognition and machine support \citep{wef2020futurejobs, musk2019neuralink}. We argue that these developments render the classical control 
problem tractable only through a cognitive co-regulation  framework---one that treats boundary-setting, epistemic  oversight, and mutual constraint as the operative mechanisms of alignment, rather than objective specification or capability limitation alone. Operationally, symbiotic coupling may be indexed by mutual information between human and AI decision trajectories over repeated interaction, or by the degree to which human epistemic outputs diverge from AI suggestions under matched conditions.

Empirical evidence suggests that AI enhances productivity when it complements rather than replaces human judgment, but only when workers retain the critical discernment to know where AI assistance helps versus where it falls short \citep{brynjolfsson2025generative, doshihauser2024creativity, dell2026navigating}. As a result, human--AI interaction is shifting toward more continuous, personalized, and cognitively intimate forms of collaboration. However, existing paradigms that frame AI either as an instrumental tool or as a social agent remain insufficient for capturing the complexity of these interactions. These models struggle to account for reflexive dynamics such as over-reliance \citep{klingbeil2024trust}, anthropomorphization \citep{folk2025individual}, and miscalibrated trust \citep{harbarth2025over}, which may ultimately exacerbate coordination failures rather than resolve them. In response, recent scholarship has begun to explore the notion of AI as becoming part of the human self---conceptualizing AI as a cognitively integrated system that co-adapts with human users \citep{collins2025artificial, noller20254e}. This perspective reframes the control problem as one of cognitive coordination and shared regulation, rather than external constraint alone.

The AI-as-part-of-self framework conceptualizes AI not simply as a separate agent but as an extension of human cognition \citep{clark1998extended, clark2025extending, hamilton2019human}. Grounded in cognitive control theory and distributed cognition, this framework argues that AI--human fusion is most effective and safest when human evaluative judgment and machine-scale computation occupy genuinely complementary roles: each doing what the other cannot, within a shared cognitive system. Cognitive control theory accounts for how humans actively maintain goals and regulate attention under uncertainty \citep{miller2001integrative, posner1990attention}. Distributed cognition holds that intelligent behavior emerges from the dynamic coupling of minds with external systems \citep{clark1998extended, hutchins1995cognition}. Together, these frameworks ground our argument that AI--human fusion is most effective when human evaluative judgment and machine-scale computation occupy genuinely complementary roles within a shared cognitive system. Rather than replacing agency, AI systems can scaffold attention, reasoning, and decision-making while preserving human oversight. However, this integration also introduces the risk of \emph{cognitive surrender}---the unreflective delegation of epistemic authority to automated systems \citep{parasuraman2000automation}. Properly structured co-regulation may mitigate alignment risks by maintaining active human control while leveraging computational augmentation \citep{schneider2019artificial, schneider2016mindscan}. 

In what follows, we examine several key areas where our AI as part of self framework applies: we explore symbiotic cognition between human and AI systems, and delegation of epistemic authority.

% ============================================================
\section{Symbiotic Cognition}
% ============================================================
\subsection{Expanding Models of the Human-AI Dyad}
A growing body of interdisciplinary research points towards a model of \emph{symbiotic cognitive labor}, in which humans and AI specialize in complementary epistemic roles while remaining mutually constrained through coordination mechanisms, boundary-setting, and oversight \citep{raisch2021automation, akata2020research}. We introduce the term \emph{symbiotic cognition} to describe a mode of cognitive organization in which humans and AI form a coupled epistemic unit with reciprocal constraint, distinct from \emph{distributed cognition} (which operates at the supra-personal level) \citep{hutchins1995cognition} and \emph{extended cognition} (which involves artifact-extension of human minds) \citep{clark1998extended}.

According to the paradigm of symbiotic cognitive labor, humans, on the one hand,
retain responsibility for goal-setting, normative judgment, contextual interpretation, and value-sensitive decisions \citep{parasuraman2000automation, sloman1996empirical}. On the other hand, AI systems contribute through large-scale pattern recognition, probabilistic inference, rapid information integration, and counterfactual exploration \citep{puranam2021humanai, dellermann2019taxonomy}. Empirical studies suggest that such designs outperform both unstructured augmentation and full automation, given the degree of human-AI complementarity present in the task \citep{fugener2025roles, vaccaro2024meta}. This perspective of AI-human symbiosis aligns with classic theories of joint cognitive systems, which argue that cognition is distributed across agents and artifacts \citep{hollnagel2006joint, hutchins1995cognition, clark1998extended}. It only functions effectively when coordination structures are explicitly designed \citep{hollnagel2006joint}. Recent work contextualizes this view under the ``distributional'' AGI hypothesis, which posits that AGI may not emerge as a single, sovereign monolithic agent, but as a modular ecosystem of specialized sub-AGI components \citep{tomasev2025distributional}. This systemic integration further contextualizes human–AI symbiosis as a collective phenomenon: the concept of the Digital We \citep{riva2025digitalwe} suggests that these AI-mediated environments increasingly shape shared identity and coordination at the group level.

In this view, symbiotic cognition is not a dyadic exchange between a person and a "black box," but an institutionalized coordination problem, where the design of the Digital We determines the quality of collective agency. Crucial to this is navigating the comfort–growth paradox \citep{riva2025digitalwe}: while AI systems optimize for psychological comfort and reduced effort, over-reliance on these frictionless interactions may inhibit the cognitive and social growth necessary for robust human–AI systems. %Such coordination in human--AI systems, we believe, requires boundary-setting mechanisms that prevent epistemic overreach by either party. By moving away from both the ``monolithic'' AI paradigm and AI-as-social-agent model, we can design for evaluative frameworks that treat the symbiotic unit as a whole, allowing the detection and mitigation for harms as they are more likely to arise in increasingly integrated systems with novel emergent capabilities and risks \citep{tomasev2025distributional}. 
Instead of epistemic centralization, such as implied by the emerging ``AI oracle'' paradigm, symbiotic cognition preserves pluralism: humans remain epistemic agents who integrate AI contributions with domain expertise, social norms, and lived
experience. However, it's important to highlight that in collectively-constituted selfhood traditions \citep{Markus1991}, the boundary between self and AI-self may be drawn differently, with implications for how epistemic authority is attributed and contested.
 The goal is not to eliminate human cognitive labor, but to reallocate it toward functions that remain difficult to automate, irreducibly human or metacognitive---judgment, responsibility, meaning-making and self-reflection---while leveraging AI as a novel cognitive function rather than a tool or a social agent \citep{raisch2021automation, puranam2021humanai,chiriatti2024system0}. 

We appropriate \emph{symbiotic cognition} \citep{kirchhoff2020symbiotic} to the human–AI context: a coupled epistemic unit of human and AI characterized by reciprocal constraint and complementary cognitive  labor. From a cognitive science perspective, symbiotic cognition aligns with extended and distributed mind theories, which argue that cognition is not bounded by the brain but is realized dynamically through interaction with external systems \citep{clark1998extended, clark2025extending}. In the context of human--AI interaction, symbiotic cognition reframes safety and alignment not as problems of domination or substitution, but as challenges of maintaining stable coordination between heterogeneous cognitive agents operating at different temporal and computational scales. A full taxonomy of cognitive terms used in this paper, situating \emph{symbiotic cognition} relative to established and emerging constructs, is provided in Appendix~A (Table 1).

\subsection{Risks: Deskilling, Cognitive Atrophy, and the Evolution of Knowledge Practices}
Concerns about deskilling are not unique to artificial intelligence. Throughout human history, cognitive practices have evolved in response to external supports, from writing and numeracy to mechanical automation and digital computation. Classic accounts of distributed and extended cognition emphasize that humans have always relied on artifacts to offload memory, calculation, and coordination, often leading not to cognitive decline but to reorganization and specialization of mental labor \citep{clark1998extended, hutchins1995cognition}. From this perspective, external cognitive scaffolds are not inherently detrimental; they are constitutive of human intelligence as it has evolved across cultural and technological environments.

Contemporary AI systems, however, introduce a qualitatively different challenge. Unlike earlier tools that supported discrete cognitive functions, generative AI increasingly performs \emph{epistemic labor} itself: generating explanations, synthesizing evidence, evaluating alternatives, and recommending actions \citep{feuerriegel2024generative}. When such systems are perceived as authoritative sources rather than assistive instruments, there is a risk of \emph{cognitive atrophy}---not merely in low-level skills, but in higher-order capacities such as critical evaluation, source monitoring, and independent judgment \citep{nickerson1998confirmation, benjamin2019assessing, lee2025impact}. This shift raises concerns that epistemic agency may erode when humans habitually defer sense-making and evaluation to opaque computational systems. This suggests the emergence of a hollowed mind, a state where the human agent maintains the outward appearance of productivity while losing the underlying "Cognitive Sovereignty" required to govern their own tools \citep{klein2025extended}. Unlike previous technologies that automated procedural tasks, AI automates the integrative reasoning that serves as the foundation for expertise, potentially leading to a permanent "Sovereignty Trap" where the costs of reclaiming independent judgment become prohibitive \citep{klein2025extended}.

Sociological studies of knowledge and expertise provide a useful lens for understanding these risks. Expertise is not simply the possession of information, but the capacity to assess credibility, contextual relevance, and uncertainty through socially learned practices \citep{collins2007rethinking}. Historically, technological systems redistributed epistemic labor while preserving human responsibility for interpretation and validation. By contrast, AI systems may collapse multiple stages of knowledge production---search, filtering, synthesis, and presentation---into a single interface, enabling uncritical acceptance and diminishing opportunities for epistemic calibration. Studies of automation bias and complacency show that when humans rely excessively on automated recommendations, their ability to detect errors and intervene appropriately declines \citep{parasuraman2000automation, skitka1999automation}. In AI-mediated environments, this dynamic may be amplified by persuasive linguistic fluency and anthropomorphic framing, which can increase perceived authority without the corresponding transparency or accountability \citep{cohn2024believing, shanahan2024talking}.

Importantly, we argue that the risk is not deskilling per se, but the \emph{unstructured delegation} that leads to deskilling. Cognitive evolution has historically involved shifts toward abstraction, meta-reasoning, and institutionalized knowledge practices. The danger arises when AI systems absorb evaluative and normative functions without reciprocal adaptation of human roles or new practices that preserve epistemic agency. Without explicit boundary-setting---such as requirements for justification, uncertainty signaling, and source diversity---AI-mediated cognition may favor convenience over competence \citep{raisch2021automation, puranam2021humanai}. From this perspective, the evolution of knowledge practices under AI should not be framed as a binary choice between preservation and replacement. Instead, it demands a deliberate redesign of the human--AI interaction to ensure that cognitive offloading supports learning, judgment, and resilience rather than undermining them \citep{tsim2025scan}. The challenge is, therefore, not to prevent cognitive change, but to guide it toward forms of symbiotic cognition that retain human responsibility for meaning, values, and epistemic accountability.

% ============================================================
\section{Delegation of Epistemic Authority}
% ============================================================

Epistemic trust---the propensity to accept information as true or action-guiding---is not a single judgment but a composite of credibility heuristics, social-normative cues, and metacognitive monitoring of sources \citep{sperber2010epistemic, metzger2007making, johnson1993sourcemonitoring}. In human communication, credibility judgments are routinely shaped by perceived expertise, benevolence, and integrity, and emerging based on inferring from the context \citep{misyak2016instantaneous}; in digital environments, they are additionally shaped by interface signals (design polish, ranking, popularity metrics), platform norms, and cues that imply institutional vetting \citep{metzger2007credibility, sundar2008main, fogg2003credibility}. These judgments function as adaptive heuristics under constraints of time and attention.

One of the key cognitive mechanisms underlying credibility is \emph{source monitoring}: the set of processes by which people remember and evaluate the origins of information (e.g., who said it, where it came from, whether it was inferred, observed, or suggested) \citep{johnson1993sourcemonitoring}. Source monitoring is deeply tied to epistemic agency: it enables individuals to attribute claims to accountable origins, triangulate across sources, and discount unreliable testimony. Complementing this cognitive account, the notion of \emph{epistemic vigilance} emphasizes that humans have evolved social-cognitive mechanisms for evaluating communicated information, including sensitivity to speaker's competence, incentives, and potential deception \citep{sperber2010epistemic}.

Recent work by Chiriatti et al.\ \citep{chiriatti2024system0, chiriatti2025system} introduces a compelling extension to dual-process theories of cognition by proposing \emph{System~0} as a distinct mode of thinking emerging from sustained human--AI interaction. Building on the classical distinction between fast, automatic \emph{System~1} processes and slow, deliberative \emph{System~2} reasoning, the authors argue that AI-mediated cognition increasingly operates \emph{prior to} and \emph{beneath} conscious human deliberation. Acting as a cognitive preprocessor, System~0 refers to pre-attentive, infrastructural cognitive activity which continuously shape the informational environment, filter possibilities, and scaffold decisions before conscious awareness is engaged. An instantiation of such System~0 dynamics can be observed in reinforcement learning–based recommendation systems. These systems operate as parallel predictive architectures: they continuously infer and update latent preferences, and selectively present information in ways that shape future cognition. In this sense, they mirror key aspects of predictive processing accounts of the human mind \citep{bargil2025}. Already, such algorithmic systems actively participate in constructing human intent and shaping perception prior to conscious awareness.
%Importantly, System~0 thinking does not possess meaning-making ability; it flags features based on processing data. As its operations are largely invisible and pre-reflective, misalignment at the System~0 level may bypass established safeguards based on conscious oversight or user consent. From a co-regulation perspective, this underscores the need for design principles that render System~0 processes legible, contestable, and adjustable. Treating AI Systems as System~0 cognition in human--AI interactions, we think, initiates the argument that alignment must address not only outcomes and decisions, but the deeper cognitive infrastructures through which human agency is enacted. 

In functional human teams, high performance is sustained through collective attention—where shared language and temporal alignment synchronize experience—and shared mental models, which enable coordinated prediction and interpretation \citep{hutchins1995cognition, salasshared2005, clark1996using}. In human–AI systems, this synchronization introduces the risk of cognitive spillover. As AI increasingly structures the informational environment by operating in System~0-like ways, it not only assists tasks but channels attention toward specific frames and strategies \citep{kahneman2011thinking, sundar2008machine, green2020dangers}. Without transparency, collective attention becomes decoupled: the human retains nominal control, yet the operative mental model is implicitly shaped by the AI. Consequently, selfhood becomes partially fragmented, as attention and model formation are externally guided. For AI to function as a genuine “extension of self,” systems must support co-attention, enabling users to access and align with the AI’s representational focus \citep{endsley1995toward, amershi2019guidelines}. In this light, shared mental models redefine human–AI complementarity beyond the human-in-the-loop paradigm toward epistemic integration, a trajectory likely to intensify with personalized, continuously adapting AI systems \citep{bansal2021does}.

Further, recent decision-theoretic frameworks suggest that the success of such human--AI symbiotic cognition depends on minimizing two distinct types of error: reliance loss (the overall frequency of delegation) and discrimination loss (the ability to identify specific instances where the AI outperforms the human) \citep{guo2024decision}. Even when humans delegate at the correct frequency, their inability to accurately differentiate signals at the 'System 0' level can lead to significant performance degradation compared to a theoretically optimal rational benchmark. While System~0 remains a theoretical proposal awaiting direct empirical validation, it provides a useful framing for understanding how AI increasingly operates as a cognitive preprocessor---structuring the informational environment, filtering possibilities, and scaffolding decisions before conscious deliberation is engaged \citep{chiriatti2024system0, chiriatti2025system}. The normative implication is not that trust in AI is inherently misguided, but that it must be structured. If source monitoring and authority assessment are offloaded to models, then systems should be designed to \emph{support epistemic agency}: exposing provenance when possible, communicating uncertainty, enabling comparison across sources, and prompting users to verify when stakes are high. Otherwise, human--AI systems risk producing a form of epistemic dependency in which the user's credibility judgments are progressively outsourced to a non-transparent mechanism. In that regime, errors are not merely factual, organizational or ethical; they are also personal, as responsibility becomes difficult to locate and contest, necessitating a paradigm shift.

\subsection{From Search Engines to Oracles}

For most of the digital era, information seeking has been a \emph{distributed epistemic activity}: users queried search engines, compared heterogeneous sources, and actively navigated uncertainty through triangulation. Search interfaces exposed multiplicity by default---ranked lists, competing viewpoints, and visible provenance---implicitly encouraging users to perform epistemic work such as source comparison and credibility assessment \citep{marchionini2006exploratory, metzger2007credibility}. Although imperfect, this model preserved a degree of pluralism and user agency in knowledge formation. The emergence of large language models and conversational AI systems marks a qualitative shift toward \emph{epistemic centralization}. Rather than returning a space of sources, these systems increasingly function as general-purpose \emph{oracles}: they aggregate information from the environment, filter it through learned representations, and present a unified synthesis as a single authoritative response. From the user's perspective, search, evaluation, and synthesis collapse into one interactional step. Although this transformation reduces friction and cognitive load, it obscures the processes by which evidence is selected, weighted, or omitted \citep{shanahan2024talking, morris2024levelsagi}.

Generative AI systems create a novel epistemic condition: they can compress search, filtering, synthesis, and explanation into a single fluent output \citep{feuerriegel2024generative}. When users treat the model as an epistemic agent---or worse, as an oracle---authority assessment can be offloaded from the user to the system. This is not simply ``cognitive offloading'' of memory; it is, we argue, a transfer of \emph{epistemic agency}. The risk is that the user retains the \emph{illusion} of understanding (because the explanation is always present), while losing the \emph{ability to audit} (because provenance is opaque or absent). Classic research on automation bias and complacency predicts precisely this pattern: when an automated system produces confident recommendations, humans tend to over-accept them and reduce independent checking, especially under time pressure or cognitive load \citep{parasuraman2000automation, skitka1999automation, skulmowski2023cognitive}.
%\subsection{Collective attention and shared mental models}

\subsection{Empirical Signals of Cognitive Reconfiguration in Human--AI Interaction}

Real-world epistemic integration is already observable: Spotify Wrapped repackages listening behavior into identity summaries that users accept as self-knowledge \citep{annabell2024spotify}; mood-tracking apps redefine emotional states through biomarker data \citep{schueller2021moodtracking}; therapeutic chatbots such as Woebot delegate introspection to algorithmic feedback \citep{darcy2023woebot}; and predictive text homogenizes self-expression by substituting algorithmic inference for intentional authorship \citep{joseph2025algorithmic}.

%Recent empirical, organizational, and policy-oriented research provides converging evidence that human--AI interaction is already reshaping cognitive practices, skill formation, and epistemic norms. Large-scale studies demonstrate that users increasingly treat generative AI systems not only as tools, but as cognitive collaborators whose outputs influence framing, persuasion, and evaluative judgment \citep{naturemachineintelligence2025interaction}. These findings suggest that the shift toward centralized AI-mediated cognition is not speculative, but already underway in everyday professional and knowledge-intensive contexts.

At the organizational level, evidence complicates the narrative of unambiguous productivity gains. While AI adoption can improve short-term efficiency, it introduces hidden penalties, including reduced perceived competence, trust calibration challenges, and shifts in responsibility attribution \citep{hbr2025hiddenpenalty, greevink2023ai, kobis2025delegation}. These dynamics suggest that even when outcomes improve, unstructured delegation may still erode epistemic agency. Policy analyses, therefore, emphasize that human advantage in AI-augmented contexts lies in higher-order capacities such as judgment, synthesis, and contextual reasoning \citep{wef2025humanskills}.

Finally, experimental work reveals that public attitudes are more nuanced than simple AI aversion. Evidence suggests that people often exhibit \emph{human favoritism}: valuing human-authored content more highly when its source is disclosed, yet do not penalize content for AI involvement \citep{jia2024humanfavoritism}. This asymmetry opens space for collaboration: when the design of human-AI interaction accounts for cognitive constraints and allocates tasks strategically to each strengths \citep{rastogi2022deciding}, the combination outperforms either alone \citep{vaccaro2024meta}---precisely the logic underlying this paper's model of symbiotic cognitive labor \citep{dellermann2019taxonomy}.

%Taken together, the empirical record confirms that cognitive reconfiguration under AI integration is not speculative but already measurable---and that its risks and benefits are distributed unevenly across individuals, organizations, and epistemic contexts.

% ============================================================
\subsection{The Impact of AI on Self-Perception and Agency}
% ============================================================

Across clinical and applied domains, closed-loop human–AI cognitive integration is already operationally real—not a theoretical projection. Neurofeedback and BCI-based interventions demonstrate that neural states can be detected, modulated, and fed back in real time to support attentional control and self-regulation \citep{sitaram2017neurofeedback, arns2009adhd}. This architecture is, in structure, a form of co-regulation. Resilience-oriented systems extend this logic further: just-in-time adaptive intervention (JITAI) frameworks deliver context-sensitive support by continuously estimating cognitive state and matching intervention to receptivity \citep{nahumshani2018jitai, mohr2014bitmodel}—a decision architecture that is, in structure, a form of co-regulation. Personalized AI assistants extend the same principle into everyday cognition: when calibrated to inferred workload, they function not as tools but as dynamic coordination layers—regulating interaction cost and support timing to preserve human agency \citep{kocaballi2019personalization, laranjo2018chatbots}. The boundary between human cognitive function and external system is permeable by design—and the question is no longer whether co-regulation occurs, but whether it is structured to preserve epistemic agency or allowed to erode it by default.

This permeability extends even to our bodily and creative self-perception. Research in dance and robotics has illustrated how AI co-presence reshapes performers' bodily self-perception \citep{troughton2026touch}. Spontaneous AI-driven variations enhance creative possibilities, blurring the distinction between dancer and machine \citep{wallace2025can}. Audience responses vary---some view the robot as an autonomous performer, while others see it as an extension of the self \citep{sathya2025cybernetic}---raising philosophical and ethical questions about authorship that mirror broader concerns about cognitive integration explored throughout this paper.

Ultimately, contemporary AI systems no longer fit within the traditional categories of \emph{tools} \citep{lake2017building} or \emph{autonomous agents} \citep{collins2024building}. Their growing role in reasoning, decision support, and creative expression suggests a paradigm shift toward \emph{cognitive co-regulation}, in which humans and AI dynamically shape each other's cognitive processes \citep{gonzalez2025cohumain, hernandez2019ai}. By framing alignment as control over an external system, we risk overlooking the fact that when AI structures what information is encountered, how alternatives are framed, and when decisions are prompted \citep{chiriatti2025system}, misalignment becomes a property of the \emph{human--AI system as a whole}, and governance must follow.

% ============================================================
\section{Alternative Views}
% ============================================================
We note three alternative views regarding our work. First, one might argue that symbiotic cognition and cognitive co-regulation are incompatible with traditional frameworks in cognitive science, which presuppose two agents capable of genuine goal-directness, intentionality, and mutual adaptation. Specifically, AI systems produce output through statistical processes \citep{bender2021dangers, messeri2024artificial}. However, symbiotic cognition operates at the \emph{functional} level: it examines the behavioral and cognitive consequences of human-AI interactions. This, therefore, avoids resolving the question of whether AI systems truly possess these intrinsic properties. This functional-level response has independent grounding in radical embodied cognitive science, which argues that cognition is constituted by agent--environment dynamics rather than internal representations alone \citep{chemero2009}. Chater et al.\ \citep{chater2022paradox} show that shared intentionality within humans emerges from the interaction process itself, whereby humans choose the tacit agreement formed via hypothetical negotiations. Symbiotic cognition functions similarly in human-AI interactions: cognition is realized dynamically through interactions with AI systems. Recent work corroborates that human-AI interaction alters human perceptual and cognitive patterns measurably \citep{glickman2025human}. Consequently, we maintain that the functional integration of AI into the cognitive loop warrants a co-regulatory approach, irrespective of the system's underlying computational nature.

It could further be argued that current related frameworks already accommodate what our work proposes. Recognizing transparency and reliability as genuine problems in human interaction with automation, Parasuraman et al.\ \citep{parasuraman2000automation} propose design solutions (appropriate automation levels, and calibrated trust) to maintain, and assume the recoverability of, human supervisory control. Modern Agentic AI, nonetheless, breaks this assumption at a more fundamental level that appropriate design cannot recover. As these systems operate across extended multi-step tasks, they disrupt the supervisory relationship before recovery \citep{cohen2024regulating}. Parasuraman et al's hierarchy, we argue, collapses when the "tool" begins restructuring the very cognitive strategies by which the human was supposed to supervise it. As a result, this necessitates a paradigm shift from supervisory control to co-regulation as the operative theoretical framework for understanding human-AI cognitive integration. 

The last alternative view emphasizes the \emph{structural} objection on symbiotic cognition. Genuine coordination requires symmetric shared intentionality and joint goal representation \citep{tomasello2005understanding}, as well as mutual adaptation at comparable timescales \citep{sebanz2021progress}. The adaptation asymmetry between human and AI systems reflects a structural difference between two types of coordination (human-human and human-AI), rendering an incorrect theoretical frame for the latter. Effective coordination, however, emerges regularly from asymmetric contributions in human-human interactions \citep{wolf2018joint}, and does not require symmetric predictability between humans \citep{konvalinka2010follow}. Thus, the symmetry requirement is an implicit assumption inherited from the specific contexts in which these frameworks were developed. Symbiotic cognition re-conceptualizes symmetry and asymmetry as \emph{dynamically emergent properties}: continuously calibrate through the interaction itself \citep{chater2022paradox}. Co-regulation produces bidirectional calibration operating continuously across different timescales \citep{clark2025extending}---a dynamic whereby humans and AI systems influence each other continuously in iterative feedback loops \citep{pedreschi2025human}. 

\section{Limitations}

This paper advances a theoretical framework and carries several limitations that future empirical and conceptual work must address. The \emph{AI as part of self} framework and symbiotic cognition model are grounded in converging theoretical traditions---extended mind theory, distributed cognition, and System~0 cognition---but have not been directly validated. Operationalizing constructs such as \emph{cognitive co-regulation}, \emph{epistemic authority transfer}, and \emph{symbiotic coupling} into measurable variables remains an open challenge. A strong assumption underlying the framework is that cognitive integration operates continuously and bidirectionally; in practice, this may hold only under conditions of sustained use, task type, or individual difference. Without experimental paradigms isolating these phenomena, the framework remains descriptive rather than explanatory \citep{Roozenbeek2024, messeri2024artificial}.

We treat ``AI'' as a relatively unified category, but symbiotic coupling likely varies substantially across retrieval-augmented models, task-specific automation, agentic planners, and embodied robotics---each imposing 
qualitatively different demands on trust calibration and epistemic delegation \citep{Morris2024, Tomashev2025}. Claims developed primarily with reference to conversational LLMs may not generalize to narrow deployments or agentic systems operating across extended time horizons. Future work should specify boundary conditions across AI system classes.

The theoretical foundations of this paper---extended mind theory, predictive coding, individualist epistemic agency---are predominantly products of Western, Educated, Industrialized, Rich, and Democratic (WEIRD) traditions \citep{Henrich2010}. Concepts such as cognitive sovereignty and self--other attribution may not translate to cultural contexts in which selfhood is relationally or collectively constituted \citep{Markus1991}. This risks producing governance recommendations parochial in scope. Extending the framework to non-WEIRD cognitive and cultural contexts is a necessary condition for global applicability. The framework's emphasis on cognitive integration raises concerns about differential impact: populations with limited AI literacy, or those subject to algorithmic systems without recourse, may experience epistemic dependence asymmetrically. Governance recommendations derived from this framework should explicitly account for fairness in how co-regulatory protections are distributed across users, contexts, and levels of technological access.

% ============================================================
\section{Conclusion}
% ============================================================
We have argued that the boundary between human cognition and artificial systems is not a fixed architectural feature to be defended, but a dynamically negotiated interface to be designed. This reframing has three direct consequences for AI alignment research. First, the unit of analysis must shift from the AI system to the human--AI cognitive system as a whole. Second, the failure modes that matter most---deskilling, epistemic authority transfer, and System~0-level misalignment---are invisible to frameworks that treat AI as an external agent. Third, cognitive co-regulation is a design requirement, specifiable in terms of boundary conditions, oversight mechanisms, and measurable epistemic outcomes.

Thus, \emph{AI as Part of Self} offers a principled path forward---one that preserves human epistemic agency not by resisting cognitive integration, but by designing it deliberately. The field now faces a tractable choice: extend alignment theory to the coupled system, or accept that its most consequential failures will remain outside the scope of any framework currently in use.

%The accelerating integration of AI into everyday reasoning, decision-making, and information synthesis marks a qualitative shift that dominant alignment frameworks have yet to absorb. Treating AI as an external agent whose objectives require external constraint is a reasonable starting point, but no longer a sufficient one.

%This paper has advanced three interlocking claims. First, AI must be modeled as a functional component of human cognition, not a discrete external system. Second, safety and alignment are properties of the coupled human–AI system as a whole, not of either constituent alone. Third, the operative mechanism of alignment must shift from external constraint to cognitive co-regulation—deliberate, bidirectional, and designed to preserve human epistemic agency under conditions of deep integration.

%The risks we identify—deskilling, epistemic authority transfer, oracle-style centralization, and System 0-level misalignment—are not speculative. Empirical signals already confirm cognitive reconfiguration in real-world human–AI interaction. The governance imperative follows directly: design for symbiotic cognition, not passive delegation. The goal is not to arrest cognitive evolution, but to ensure it proceeds without surrendering the epistemic foundations of human autonomy.

\begin{ack}
% Acknowledgments go here for camera-ready version.
% Do not include in anonymized submission.
\end{ack}

% ============================================================
%  References
% ============================================================
% Compile: pdflatex -> bibtex -> pdflatex -> pdflatex
% Upload refs.bib alongside this file in Overleaf.

\bibliography{refs}

\newpage
%\documentclass{article}
%\usepackage{booktabs}
%\usepackage{tabularx}
%\usepackage{ragged2e}
%\usepackage[margin=2cm]{geometry}

%\begin{document}
%\pagestyle{empty}

\section*{Appendix A: Taxonomy of Cognitive Terms}

{\small This appendix provides a structured taxonomy of the key
cognitive terms used throughout the paper, distinguishing
\textit{established} terms with settled definitions in the cognitive
science literature, \textit{emerging} terms that are gaining traction
but lack consensus definitions, and our \textit{proposed} term
\emph{symbiotic cognition}, which we appropriate and extend to the
human--AI context. $\dagger$~indicates our proposed or extended usage.}

\bigskip

\noindent\textbf{Table 1: Taxonomy of cognitive terms: established, emerging, and proposed.}

\bigskip

\noindent\renewcommand{\arraystretch}{1.3}
\scriptsize
\setlength{\tabcolsep}{4pt}
\begin{tabularx}{\textwidth}{>{\RaggedRight\arraybackslash}p{1.8cm}
                              >{\RaggedRight\arraybackslash}p{2.0cm}
                              >{\RaggedRight\arraybackslash}X
                              >{\RaggedRight\arraybackslash}p{2.8cm}}
\toprule
\textbf{Term} & \textbf{Source} & \textbf{Description} & \textbf{Example} \\
\midrule
\multicolumn{4}{l}{\textit{Established Terms}} \\
\midrule

Extended mind &
Clark \& Chalmers (1998) \citep{clark1998extended} &
Cognition extends beyond the brain into artifacts and environment;
external resources become constitutive of cognitive processes. &
Using a notebook as functional external memory. \\
\addlinespace[2pt]

Distributed cognition &
Hutchins (1995) \citep{hutchins1995cognition} &
Cognitive processes distributed across people, artifacts, and
environment at supra-personal level; the system, not the individual,
is the unit of analysis. &
Cockpit crew and instruments jointly navigating an aircraft. \\
\addlinespace[2pt]

Cognitive offloading &
Risko \& Gilbert (2016) \citep{risko2016offloading} &
Deliberate use of external resources to reduce internal cognitive
demand, preserving capacity for higher-order processing. &
Using a calendar to offload scheduling decisions. \\
\addlinespace[2pt]

Epistemic agency &
Sperber et al.\ (2010) \citep{sperber2010epistemic} &
Capacity to evaluate, monitor, and take responsibility for one's own
knowledge formation, including source assessment and credibility
judgment. &
Independently verifying a source before accepting a claim. \\
\addlinespace[2pt]

Automation bias &
Parasuraman et al.\ (2000) \citep{parasuraman2000automation} &
Tendency to over-rely on automated recommendations, reducing
independent verification under time pressure or high cognitive load. &
Accepting a flawed AI diagnosis without verification. \\
\addlinespace[2pt]

System 1 / System 2 &
Kahneman (2011) \citep{kahneman2011thinking} &
Dual-process framework: fast, automatic (System~1) vs.\ slow,
deliberative (System~2) cognitive processing. &
Intuitive vs.\ analytical response to a risk decision. \\

\midrule
\multicolumn{4}{l}{\textit{Emerging Terms}} \\
\midrule

System 0 &
Chiriatti et al.\ (2024, 2025) \citep{chiriatti2024system0, chiriatti2025system} &
Pre-attentive, AI-mediated cognition operating prior to conscious
deliberation; AI structures the informational environment before
awareness is engaged. &
AI news feed shaping belief formation before conscious reading begins. \\
\addlinespace[2pt]

System~3 (Tri-System Theory) &
Shaw \& Nave (2026) \citep{shaw2026system3} &
Extends dual-process theory by positing System~3: artificial cognition
operating outside the brain that can supplement or supplant System~1
(intuition) and System~2 (deliberation). &
A knowledge worker submits AI output verbatim, overriding both
intuitive doubt and deliberate re-reading. \\
\addlinespace[2pt]

Complementary Intelligence &
Gonzalez \& Malloy (2026) \citep{gonzalez2026complementary} &
AI complements rather than replicates human cognition: cognitive AI
models human perception and decision-making; machine AI handles
data-driven optimization at scale. &
AI handles pattern detection; clinician handles judgment and
ethical reasoning. \\
\addlinespace[2pt]

Bidirectional alignment &
Li \& Song (2025) \citep{li2025co} &
Mutual cognitive adaptation between human and AI systems across
interaction; alignment as a two-way calibration process. &
User and AI co-adapting decision strategies over repeated use. \\
\addlinespace[2pt]

Cognitive sovereignty &
Klein \& Klein (2025) \citep{klein2025extended} &
Individual's capacity to govern their own cognitive tools and resist
epistemic dependence. &
Retaining independent judgment despite external systems (e.g., AI) recommendations. \\
\addlinespace[2pt]

Epistemic authority transfer &
Parasuraman et al.\ (2000) \citep{parasuraman2000automation} &
Gradual delegation of credibility assessment to an automated system,
reducing the user's capacity to audit knowledge claims. &
Trusting AI-generated summaries without auditing sources. \\
\addlinespace[2pt]

Cognitive co-regulation &
Shen et al.\ (2025) \citep{shen2025bidirectional} &
Dynamic mutual shaping of cognitive processes between human and AI
within a shared system; neither party fully controls the epistemic
outcome. &
Human and AI jointly calibrating decision thresholds in real time. \\

\midrule
\multicolumn{4}{l}{\textit{Our Proposed Term}} \\
\midrule

Symbiotic cognition$^\dagger$ &
Extended from Kirchhoff (2020) \citep{kirchhoff2020symbiotic} &
A coupled epistemic unit of human and AI characterized by reciprocal
constraint and complementary cognitive labor --- distinct from
\textit{distributed cognition} (supra-personal level) and
\textit{extended cognition} (artifact-extension of individual minds). &
Human retains normative judgment; AI provides probabilistic
inference; neither operates independently of the other's constraints. \\

\bottomrule
\end{tabularx}

\bigskip

\noindent{\small\textbf{Note.} We exclude \textit{swarm intelligence}
and \textit{liquid brains} as these operate at collective,
decentralized scales incompatible with the dyadic human--AI system
this paper addresses.}

%\end{document}

\end{document}